\documentstyle[prl,aps,multicol,graphics]{revtex}

\begin{document}
\draft

\title{Statistical Distribution of Intensities Reflected from  Disordered
Media}

\author{A. Garc\'{\i}a-Mart\'{\i}n, T. L\'{o}pez-Ciudad,
J.J. S\'{a}enz}

\address{Departamento de F\'{\i}sica de la Materia Condensada
and Instituto de Ciencia de Materiales ``Nicol\'{a}s Cabrera'',
Universidad Aut\'{o}noma de Madrid, E-28049 Madrid, Spain.}

\author{M. Nieto-Vesperinas}

\address{Instituto de Ciencia de Materiales de Madrid.  Consejo Superior
de Investigaciones Cient\'{\i}ficas, Campus de Cantoblanco, E-28049 Madrid,
Spain. }

\date{12 March 1998}
\maketitle
\begin{center} {\bf Phys. Rev. Lett. 81, 329 (1998)}\\ \copyright 1998
The American Physical Society \end{center}
\begin{abstract}

A theoretical analysis of the statistical distributions of the reflected
intensities from random media is presented.  We use random matrix 
theory to analytically deduce the probability
densities in the localization regime.  Numerical calculations of the
coupling to backward modes in surface corrugated waveguides are also put
forward for comparison.  Interestingly, the speckle distributions are
found to be independent of the transport regime.  Despite the scattering
being highly non-isotropic, the predicted probability densities
reproduce accurately the numerical results. \\

\end{abstract}
\begin{multicols}{2}
%
%

The scattering of both classical and quantum waves
from random media   has long been a subject of
interest \cite{mogo1}.
Interesting multiple scattering effects like
the enhanced  backscattering (EB) \cite{ecb,ecbs} and intensity
correlations in transmitted and reflected waves
\cite{porron,Mello1,Tit1}
have extensively been analyzed.
Recently
much attention has been focused on this subject in
connection with the probability distributions of the different
transmittances in disordered waveguides 
\cite{talytal,Bee,experiments,Jator,tobeF}.
Rayleigh and Gaussian statistics with anomalous tails
in the diffusive regime, observed experimentally\cite{experiments},
were analytically deduced
within the random matrix theory (RMT) framework
\cite{talytal,Bee-rev}. A strong dependence on the
length $L$ of the disordered region was expected from numerical
simulations of waveguides with volume disorder \cite{Edrei}.
Recent results\cite{Bee}, also based on RMT, have shown
how the Rayleigh and Gaussian
distributions  evolve into
the same lognormal distribution as $L$
increases beyond the localization length $\xi$.
The same behavior has been found in simulations of surface
corrugated waveguides \cite{tobeF}.
However, in contrast with the transmission
case, the probability distributions of reflected intensities have not
been discussed in detail previously.
It is the purpose of this work to
put forward a comprehensive picture of the
statistics of reflected waves from a
disordered waveguide based on analytical as well as on
numerical calculations.

In order to discuss some general properties of the backward scattering,
random matrix theory provides a useful macroscopic approach since it
does not depend on the specific details of the scattering processes.
We will make use of RMT results to
analytically deduce the probability
density of the reflected {\em speckle pattern} in a $N$-mode
waveguide in the localized regime. As expected
from the central limit theorem,
in the limit of large $N$, the speckle distribution follows a negative
exponential law (Rayleigh statistics).
As a consequence of enhanced coherent
backscattering effects (i.e. time-reversal invariance \cite{Tit1}),
as $N$ decreases, this Rayleigh law evolves into two different
distributions depending on the backward scattering directions.
The numerical calculations of the speckle statistics
for a  corrugated waveguide
show a remarkable agreement with these analytical results
 all the way from the diffusive through the strong localized regime.
Although the analytical results are formally deduced in localization,
this agreement indicates that, in contrast with transmission,
{\em the  reflected speckle distributions are
 independent of the transport regime}.

Our theoretical approach is based on the properties of the 
scattering matrix $S$ of a waveguide with a
central region containig  the random medium 
(whose disorder can be indistinctly either in volume or in the walls)
(see Fig.1). Outside this region, the transverse
confinement in the waveguide defines $N$ propagating modes or channels
\cite{pedete}.
The wave transport can be described in terms of the scattering matrix $S$:
\begin{eqnarray}
\left(\begin{array}{cc} r&\tilde{t}\\t &\tilde{r}\end{array}\right)\ ,
\end{eqnarray}
where $ r,\ \tilde{t},\ t,\ \tilde{r},\ $ are square matrices whose
dimension is given
by the total number of channels $N$.
The matrix elements $r_{ij}$ and $t_{im}$ are the reflected and
transmitted amplitudes into the channels $j$ and $m$, respectively, 
 when there is a unit flux in the channel $i$ incident 
 from the left; $\tilde{r}_{nm}$
and $\tilde{t}_{nj}$ have the same meaning but the incoming channel is
incident from the right.  In terms of these elements, the transport
coefficients  \( R_{ij}=|r_{ij}|^2\) and \(
T_{im}=|t_{im}|^2\) measure the reflected and transmitted speckle
patterns respectively, being \(R_i=\sum_j R_{ij}\)
and \(T_i=\sum_m T_{im}\) the total reflection and transmission for the
incoming mode $i$.

The study of the probability densities of the
reflected intensities \( P(R_{ij}) \)
is much simpler in the strong localization
regime where the transmission amplitudes
vanish. The matrix $S$ is then reduced to the two square blocks $r$
and $\tilde{r}$.
Since we are interested in a statistical approach, given our limited
knowledge of the microscopic coupling between channels, the natural
choice of the statistical ensemble is that which maximize the
information entropy subject to the known constrains 
[ {\em flux conservation} ($S$ unitary) and
{\em time-reversal invariance} ($S$ symmetric) ].
In the RMT context, this leads to the circular orthogonal ensemble
(COE, $\beta=1$), i.e.
the phases of
the elements of the reflection blocks are as random as
possible while their moduli are constrained by the conditions of
unitarity and symmetry.
In the absence of time reversal   invariance, the
matrix $S$ loses its symmetry and the appropriate
statistical ensemble is then the circular unitary  ensemble (CUE,
$\beta=2$).

Following  general formulas for the
distribution  of matrix elements in the circular ensembles derived
by Pereyra and Mello \cite{PerMel}, we can compute the
distribution of the reflected speckle pattern
\( P(R_{ij}) \). Due to the underlying {\em isotropy} 
condition \cite{Mello1,Bee-rev} the RMT results are 
mode independent except for the  backscattering into the 
same channel:
\begin{eqnarray}
P(R_{ii},\beta=1) = \left(\frac{1}{\langle R_{ii} \rangle} -1\right)
(1-R_{ii})^{1/\langle R_{ii} \rangle -2} \ ,
\label{iibeta1}
\end{eqnarray}
and
\begin{eqnarray}
&&P(R_{ij},\beta=1) = \left(\frac{1}{2\langle R_{ij} \rangle} -1\right)
(1-R_{ij})^{1/\langle R_{ij} \rangle -3}
\nonumber\\
&&\times\ {}_2F_1\left( \frac{1}{2\langle R_{ij} \rangle} -1,1 ;
\frac{1}{\langle R_{ij} \rangle} -2, 1-R_{ij} \right) \ ,(i\neq j) \ ,
\label{ijbeta1}
\end{eqnarray}
where ${}_2F_1$ is the hypergeometric function and
the mean values $\langle R_{ij} \rangle$ are given by 
\cite{Mello1,Tit1}:
\begin{eqnarray}
\langle R_{ij} \rangle =
\frac{1+\delta_{ij}\delta_{\beta 1}}{N+\delta_{\beta 1}}\ . 
\label{promij} \end{eqnarray}
Although recent numerical calculations
\cite{tobeF,PelosPRB96,JASan} show that the different
channels are not  equivalent, as we will see,  the
actual statistical  distributions can still be described by the general
RMT results. 
The key point will be to consider the distributions as functions of the mean
values rather than in terms of the number of channels.

%
%
%
%
%


In Figs. 2 and 3 we have plotted Eqs.  (\ref{iibeta1}) and
(\ref{ijbeta1}) for a set of values of $\langle R_{ii}\rangle$ and
$\langle R_{ij}\rangle$, respectively (thick solid and broken lines),
together with  numerical results for a surface corrugated waveguide
(see the details below). To this end,
$\langle R_{ij}\rangle$ always involves $i \ne j$.
In the limit of $N$ large , 
(both $\langle R_{ii}\rangle$ and $\langle R_{ij}\rangle$ $\ll 1$),
both distributions (\ref{iibeta1}) and (\ref{ijbeta1}) 
evolve into a Rayleigh law.
Coherent backscattering effects manifest themselves
in the distributions for small $N$ which dramatically depend 
on whether $i=j$, or $i \ne j$.
This is  illustrated in the analysis of the
case $N=2$  ($\langle R_{ii}\rangle = 2/3$
and $\langle R_{ij}\rangle = 1/3$) where
the probabilities
for $i=j$ and for $i\neq j$ are quite different (see inset in
Fig. 3):
\begin{eqnarray} P(R_{ii},\beta=1) =
\frac{1}{2 \sqrt{1-R_{ii}}} \ , \label{n2ii} \end{eqnarray} and
\begin{eqnarray} P(R_{ij},\beta=1) = \frac{1}{2 \sqrt{R_{ij}}} \ , (i \ne j)\ .
\label{n2ij} \end{eqnarray}

When  time reversal symmetry is removed, 
there is no longer difference between $P(R_{ij})$ 
and $P(R_{ii})$, and hence there
is a unique intensity distribution.  Interestingly, in terms of the
averages the statistical law coincides with that obtained for $R_{ii}$
in the case of $\beta=1$, i.e.
\begin{equation}
P(R_{ij},\beta=2)=P(R_{ii},\beta=2)=P(R_{ii},\beta=1)\ .
\end{equation}
This means that the {\em statistical  law} that
governs the backscattered intensities $R_{ii}$ does not depend on  
the time reversal symmetry conditions, 
as long as they are expressed in terms of the mean value.

It is worth noticing, that
for a chaotic cavity\cite{baranger},
 Eqs. \ref{iibeta1} and \ref{ijbeta1} 
are exact
results tor the reflected intensities for any $N$.
In this
case, $\langle R_{ij} \rangle =
(1+\delta_{ij}\delta_{\beta 1})(2N+\delta_{\beta 1})^{-1}$.
Again, in terms of the mean values, the statistics  of
reflected waves from a chaotic
cavity and from a waveguide in the  localization regime are
indistinguishable.

Let us now discuss the numerical results for the
reflection
coefficients in a
surface disordered waveguide (see Fig. 1).
The corrugated part of the waveguide, of total length
$L$ and perfectly reflecting walls, is composed of $n$ slices of length
$l$. The width of each
slice has random values uniformly distributed between $W_0-\delta$ and
$W_0+\delta$ about a mean value $W_0$.  Following previous works, we
shall take $W_0/\delta= 7$ and $l/\delta = 3/2$.
The main transport properties do not depend on the particular choice of
these parameters, however\cite{tobeF,APL}.    Transmission and
reflection coefficients are exactly calculated by solving the 2D wave
equation by mode matching at each slice, together with a generalized
scattering matrix technique\cite{APL,Weiss,JosanMetodo}.
In order to obtain the intensity distributions as well as the mean values
for the reflection coefficients, we have performed the calculations over an
ensemble of one thousand  configurations.

The anisotropy of the scattering process, predicted in Refs.
\cite{tobeF,PelosPRB96,JASan}, is specially relevant when
the EB factor $\eta_i$, defined as
\begin{eqnarray}
\eta_{i}= (N-1) \frac{\langle R_{ii}\rangle}{\langle R_{i}\rangle-\langle
R_{ii}\rangle}\ ,
\end{eqnarray}
is addressed.
This EB factor is very sensitive to the changes in the mean values, 
and so it is
specially affected by changes in the incident wavelength.
In Fig. 4b we show the behavior of  $\eta$ as a function of the wavelength, for
a length of the corrugated part such that the system is always in the
localization regime.
The behavior of the ``external'' modes (defined as those propagating
modes with either the smallest or the largest transversal momentum) is
quite different to that of the ``central'' ones.  The lowest mode
(i.e. that with the smallest transversal momentum) has a marked
oscillatory value of $\eta$, showing peaks
at the onset of each new propagating mode in the waveguide.
For the rest of the modes, except for
small oscillations,
$\eta$ evolve
from a high value  just at the onset of propagation, to the expected
factor of 2. This oscillating behavior is closely related to that
observed in
the localization length $\xi$\ shown in Fig. 4a, (see also Ref. \cite{APL}).

The numerical probability densities have been calculated in the
range of wavelengths shown in Fig. 4, for different length values
including  both the diffusive ($\xi /N \lesssim L\lesssim
\xi$) and  localized    ($ L \gtrsim \xi $) regimes.  The histograms
in Fig. 2 and 3 (thin solid lines) show the intensity distributions
numerically obtained for different averages of either $\langle
R_{ii}\rangle$ or $\langle R_{ij}\rangle$.  Distributions corresponding
to different number of propagating modes and different transport regime,
but having the same average of the reflection coefficient, are
indistinguishable. Despite being the scattering highly non-isotropic,
it is remarkable that both numerical and analytical results coincide
with each other, without any adjustable parameter.
This indicates  that, while the average values depend 
on the specific system under consideration, the fluctuations, which are 
characteristics of disordered systems, are fully represented by the
probability density {\em as a funtion of the mean values}, and have general
properties that are well described by RMT.

In summary, we have derived the general statistical properties of the
reflected  intensities from a random medium in the localization regime,
based on RMT.
Comparisons have been made by numerically analizing
the coupling to backward modes in surface corrugated
waveguides. The enhanced backscattering factor can be much larger than that
predicted by RMT, and  oscillates versus the wavelength following
the  oscillating behavior observed for the localization length.
We show, however, that in spite of the anisotropy of the interactions,
the numerical statistical distributions accurately follow the RMT
predictions, independently of the transport regime.
This indicates that the statistical distributions do not depend on the
details of the scattering processes. They should then be observable
in different systems under rather general disorder conditions.

We are indebted to A.J. Caama\~{n}o and J. A. Torres 
for  stimulating discussions.
This work has been supported by the Direcci\'{o}n General de Investigaci\'{o}n
Cient\'{\i}fica y T\'{e}cnica (DGICyT) through Grant No.  PB95-0061 and  by
the  European Union. A. G.-M.
acknowledges partial financial support from the postgraduate grant
program of the Universidad Aut\'{o}noma de Madrid.

%
\begin{figure}
\narrowtext
%
\centering \resizebox{7.5cm}{!}{\rotatebox{-90}
{\includegraphics*[5.cm,0cm][14cm,27cm]{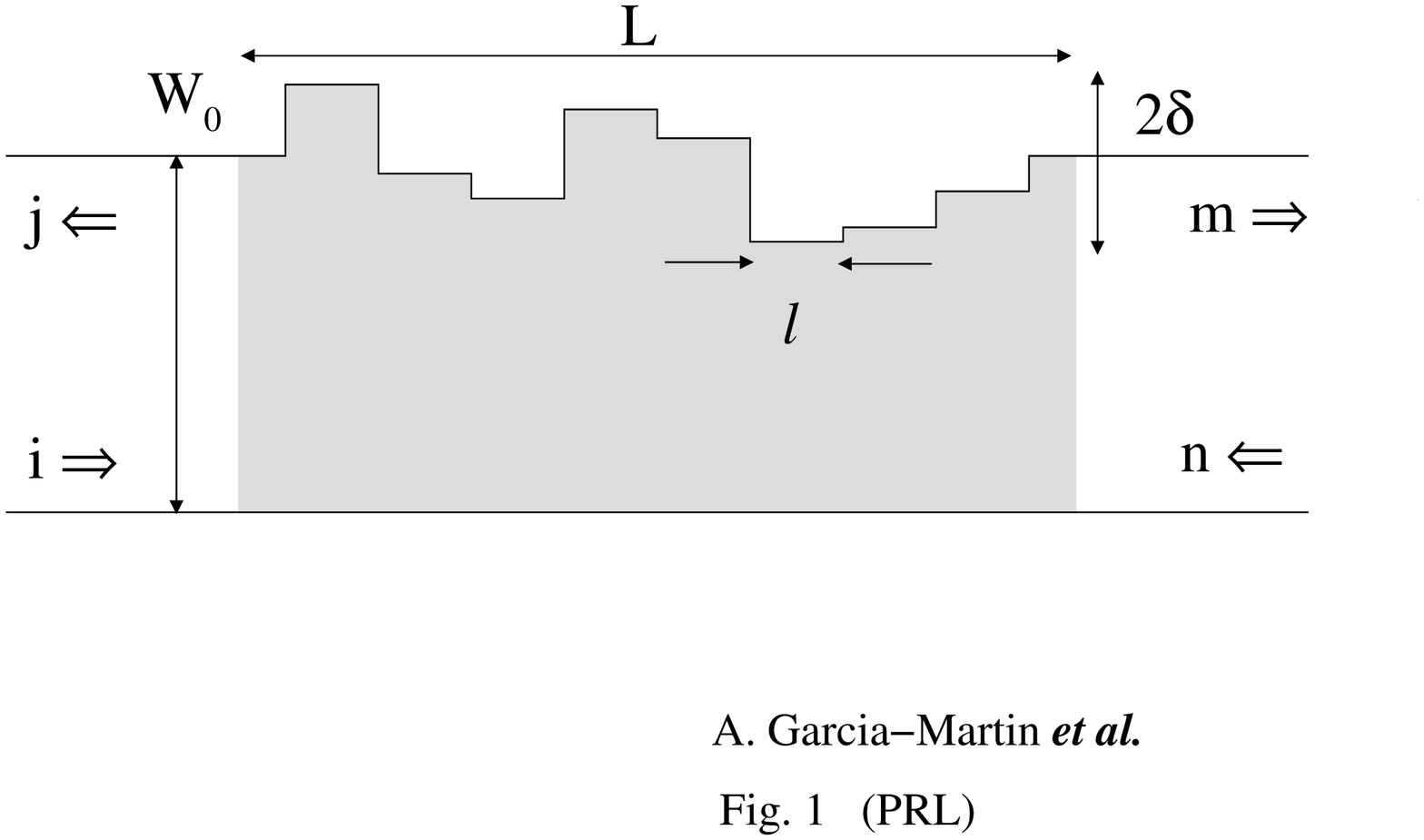}}}
\caption{
 Schematic view of the system under consideration.}
\label{Fig. 1}
\end{figure}
%
%
%
%
%
\begin{figure}
%
%
\narrowtext
%
\centering \resizebox{7.5 cm}{!}{\includegraphics*[0.75cm,5cm
][20cm,26.5cm]{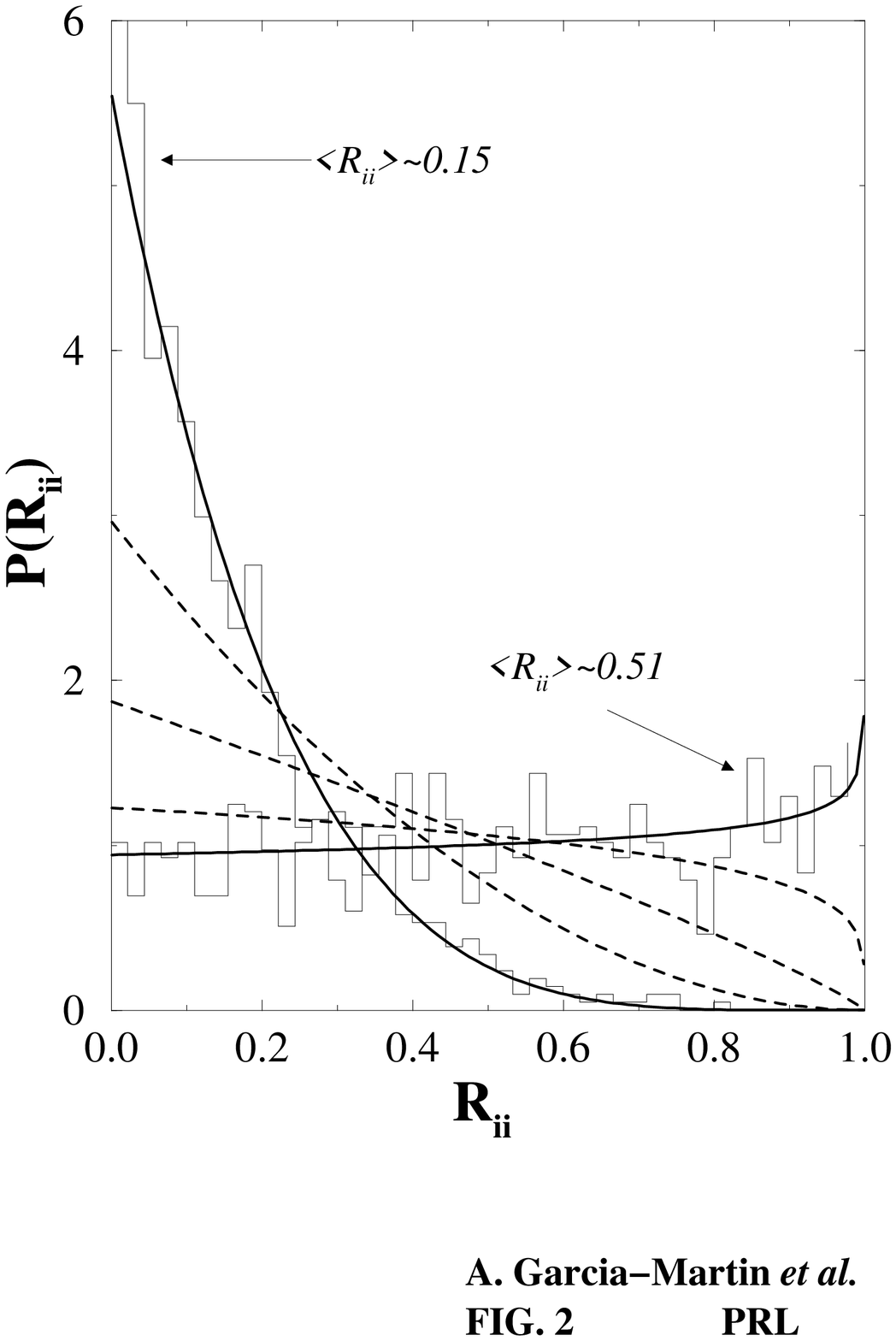}}
%
%
\caption{
Distributions of $R_{ii}$ for different
values of $\langle R_{ii} \rangle$. Thin solid lines correspond
to the numerical calculations for
$\langle R_{ii} \rangle \sim 0.15, 0.51$, respectively, whereas thick
solid lines are the analytical predictions.
Thick broken lines are the analytical results for
$\langle R_{ii} \rangle \sim 0.25, 0.35$ and $0.45$, respectively.
}

\label{Fig. 2}
\end{figure}
%
%
%
%
\begin{figure}
%
%
\narrowtext
%
\centering \resizebox{7.5cm}{!}{\includegraphics*[1cm,6cm
][20cm,25cm]{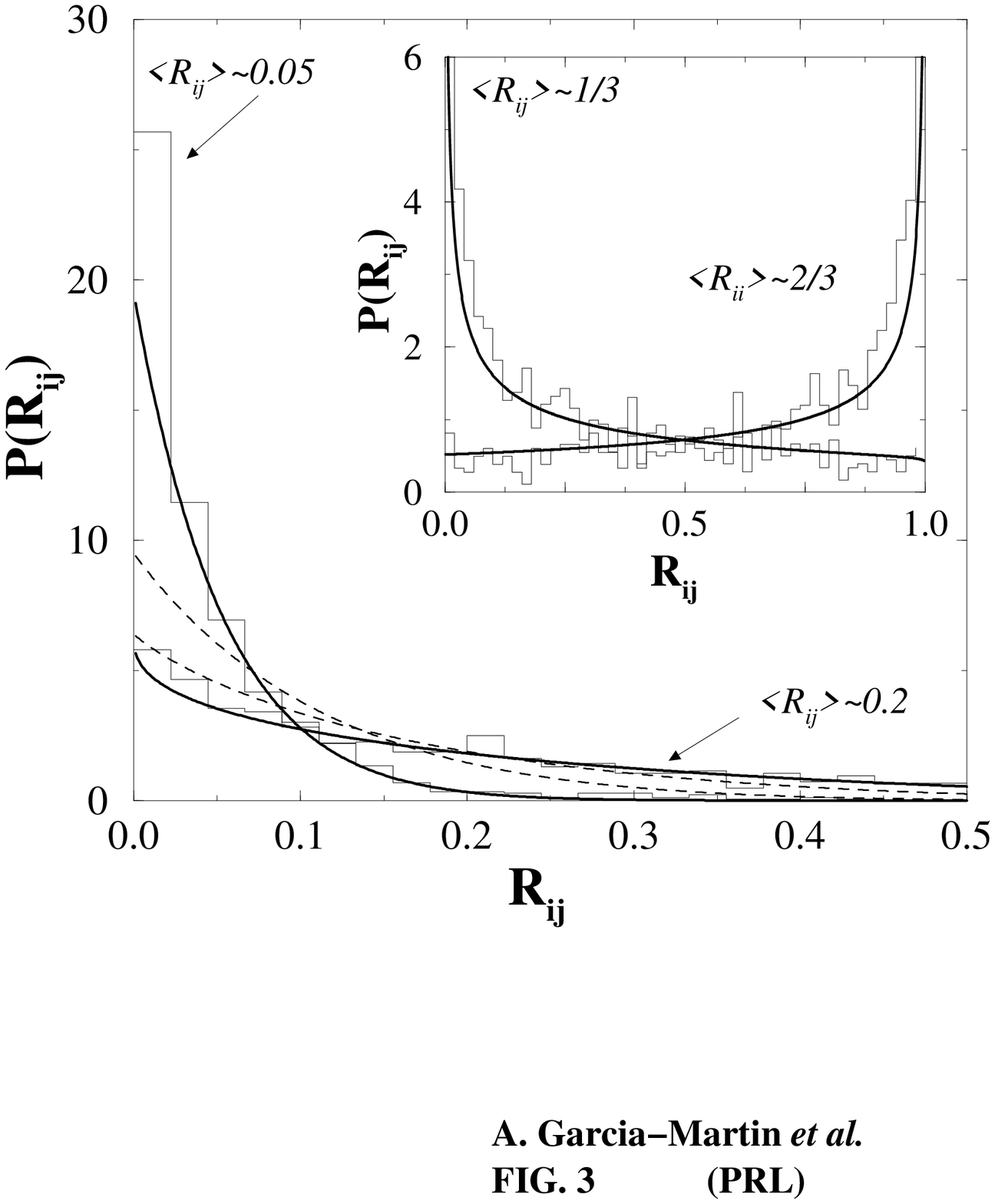}}
%
%
\caption{
Distributions of $R_{ij} (i \ne j)$ for different
values of $\langle R_{ij} \rangle$.  Thick
solid lines are the analytical predictions for the same values of
the averages, as those indicated for the numerical histograms
(thin solid lines)
($\langle R_{ij} \rangle \sim 0.05, 0.2$, respectively).
Thick broken lines are the analytical results for
$\langle R_{ij} \rangle \sim 0.10$ and $0.15$, respectively.
Inset: Probability density for $N=2$: $P(R_{ii})$ 
($\langle R_{ii}\rangle = 2/3)$, and $P(R_{ij})$ 
($\langle R_{ij}\rangle = 1/3$).}
\label{Fig. 3}
\end{figure}
%
%
%
%
%
 \begin{figure}
%
%
%
%
\narrowtext
\centering \resizebox{7.5cm}{!}{\includegraphics*[0.75cm,6cm
][20cm,26cm]{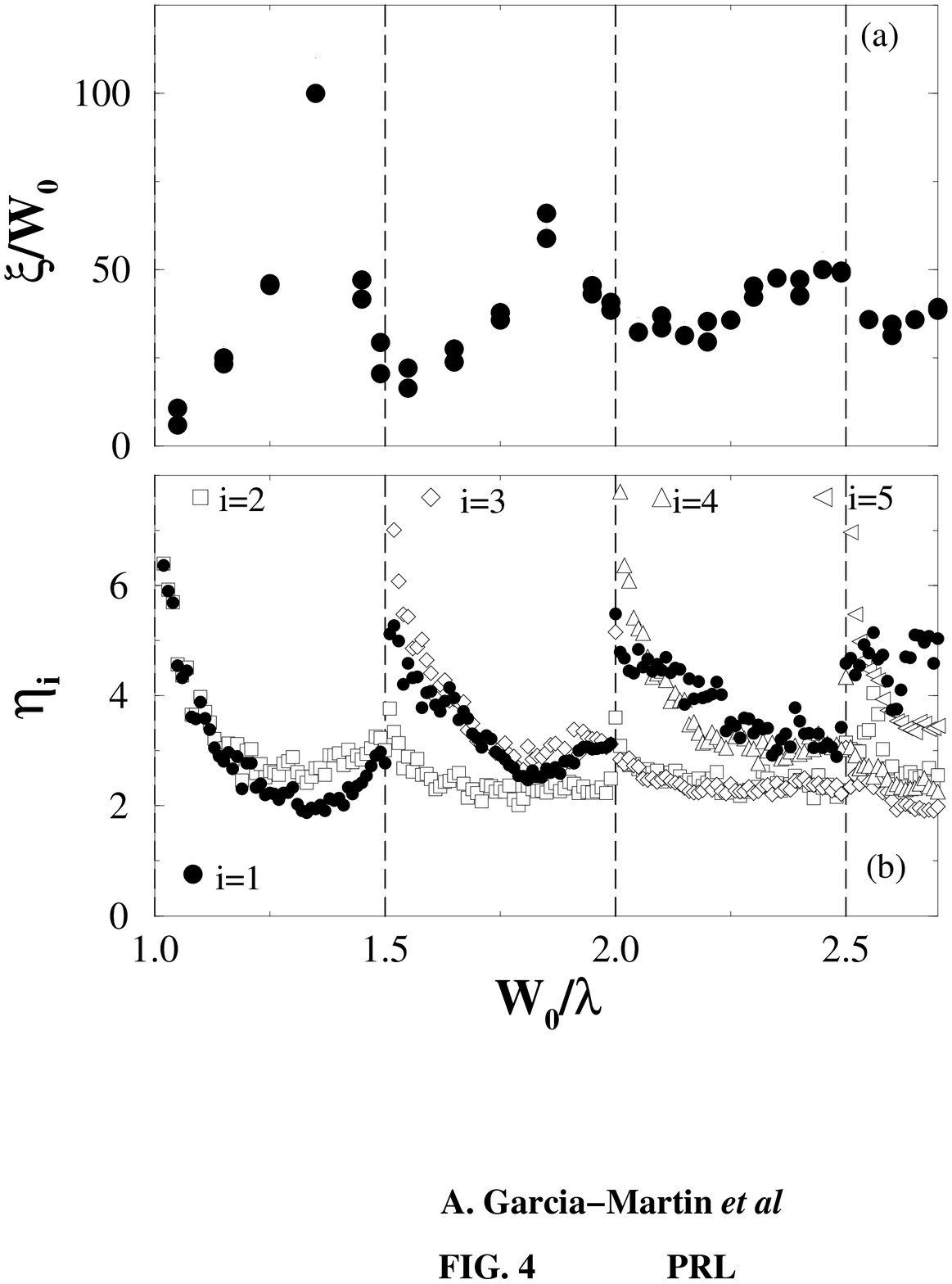}}
\caption{
(a) Localization length $\xi$ versus $W_0/\lambda$.
(b) Enhanced backscattering factor $\eta_i$ versus
 $W_0/\lambda$. Vertical long-dashed  lines are drawn
at the onset of a new mode.}
\label{Fig. 4}
\end{figure}
%
%
%
%
\end{multicols}


\begin{references}
\bibitem{pedete} In the case of a sample in
free space no
discrete modes exist. Here, a channel is a solid angle containing one
coherence area or - in practical terms - having the size of a speckle
spot. The number of channels in this case is given by $N=2 \pi
A/\lambda^2$ where $A$
is the cross section of the waveguide or the illuminated area of the
sample.
\end{references}
\end{document}